\documentclass{sf2a-conf}
\usepackage{graphicx}
\begin{document}

\TitreGlobal{SF2A 2006}

\title{Determining the evolutionary history of galaxies by astrocladistics : some results on close galaxies.}
\author{Fraix-Burnet, D.}\address{Laboratoire d'Astrophysique de Grenoble, BP 53, F-38041 Grenoble cedex 9, France}
\runningtitle{Evolutionary history of galaxies by astrocladistics}
\setcounter{page}{237}
\index{Fraix-Burnet, D.}

\maketitle
\begin{abstract} 
Astrocladistics, a methodology borrowed from biology, is an objective way of understanding galaxy diversity through
evolutionary relationships. It is based on the evolution of all the available parameters describing galaxies
and thus integrates the complexity of these objects. Through the formalization of the concepts around galaxy formation
and evolution, and the identification of the processes of diversification (build up, secular
evolution, interaction, merging/accretion, sweeping/ejection), galaxy diversity can be expected
to organize itself in a hierarchy. About 500 galaxies described by about 40 observables have now been analysed
and several robust trees found. For instance, we show that the Dwarf Galaxies of the Local Group all derive from a common ancestral
kind of objects. We identify three evolutionary groups, each one having its own characteristics and own evolution. 
The Virgo galaxies present a relatively regular diversification, with rather few violent events such as major mergers. Diversification in another sample made of gas-poor galaxies in different environments appears to be slightly more complicated with several diverging evolutionary 
groups. Work on a large sample of galaxies at non-zero redshifts is in progress and is pioneering a brand new approach to exploit data from the big extragalactic surveys. 
\end{abstract}
%
\section{Introduction}

Astrocladistics is a methodology of data analysis that groups objects according to their evolutionary history (Fraix-Burnet 2004; Fraix-Burnet et al. 2003, 2006a, 2006b, 2006c). It is particularly suited to complex objects in evolution. It is based on \emph{all} available observables. These properties are a priori considered as characterizing the evolutionary behaviour of a particular component of a galaxy, and are thus divided into several evolutionary states. Any change between two states is a progress in the evolution that is transmitted to the subsequent, hence more evolved, kinds of objects. Using all of these evolutionary properties, the cladistic analysis groups objects by estimating how much evolution is required between them. The final picture is depicted on a hierarchical tree, called a cladogram, in which the distance between two branches reflects the level of diversification that occurred during evolution. This is
unlike trees obtained from multivariate distance analyses that only show differences averaged over all properties without any consideration of the detailed evolution of the different components in a galaxy. But this has a cost because the cladistic analysis is much longer to perform. The cladistic analyses can be performed on evolutionary groups of objects rather than individuals. This is obviously the key to deal with data from big surveys, in which hundreds of thousands galaxies are described by about a hundred observables at most. Up to now, we have obtained results on still modest samples of close galaxies, and we progressively resolve the challenge of including more and more objects and combining samples described by non-matching sets of observables.

\section{Diversity of the dwarf galaxies of the Local Group}

The astrocladistics analysis of the 36 dwarf galaxies of the Local Group, described by 24 characters, successfully yields a robust tree, demonstrating that all these objects have diversified from an ancestral common kind of objects (Fraix-Burnet et al. 2006c). This result is already important because it confirms the applicability of cladistics to galaxies, and more importantly, opens the way toward determining, in the future, the nature of the very first objects of the Universe.

\begin{figure}[h]
   \centering
   \includegraphics[width=9cm]{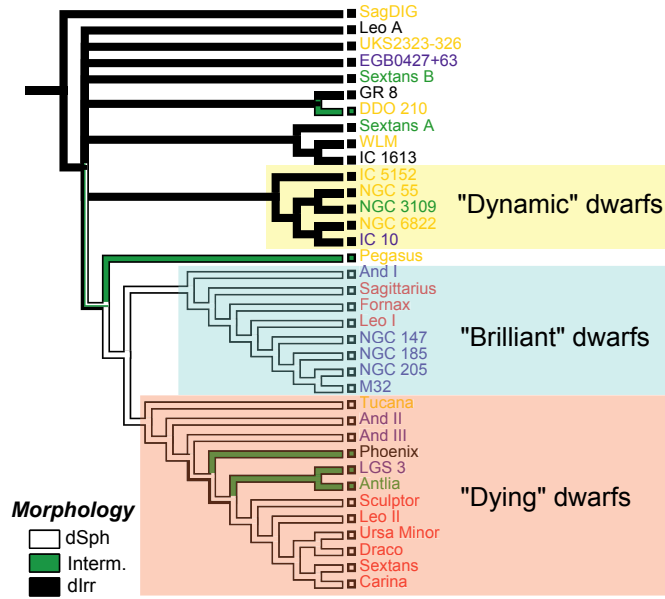}
      \caption{Cladogram for the Dwarf Galaxies of the Local Group. Three evolutionary groups are identified on the tree and characterized from their properties.}
       \label{fignaines}
   \end{figure}
   
Several astrophysical results are derived from this study. The galaxy SagDIG is the less diversified from an ancestral stage common to all this sample. Interestingly, SagDIG happens to be at the edge of the Local Group. By projecting values of the observables onto the tree, we conclude that the evolutions of the properties are rather complex in this sample of the Local Group, showing that the dwarf galaxies are certainly easily influenced by their environment. We also derive the characteristics of some groups corresponding to structures in the tree. Three are obvious and identified on Fig.~\ref{fignaines}. The group that we call ``dynamic'', gathers galaxies showing a lot of gas, strong star formation, and a complicated kinematics. The second one, called ``brilliant'', gathers galaxies which are rather luminous but with a gentle kinematics and not much star formation. The last one, the ``dying'' group, has objects which seem to be the dead-end of dwarf galaxies, with no gas and a low luminosity. These three groups are all well differentiated from SagDIG, they all evolved in their own way.

This work can be certainly improved by incorporating new data which are much more detailed, both spectroscopically and spatially. However, it is already clear from Fig.~\ref{fignaines} that the traditional morphological dichotomy is too simplistic to describe the diversity and the evolution of dwarf galaxies. We rather find that the ancestral group is of irregular shape (dIrr), but some other kinds of dIrrs appeared during evolution (the ``dynamic'' group), while there are two very distinct groups of spheroidal galaxies (dSph). The three groups that we have identified reveal a more physical insight into the origin of the diversity of these galaxies.

\section{Evolutionary history of Virgo galaxies}

In a subsequent study, Fraix-Burnet \& Davoust (2006) analysed 222 galaxies of the Virgo cluster, described by 48 observables. Again, a nicely resolved tree is found, demonstrating that all these galaxies derive from the same kind of objects. This tree depicts the probable evolutionary scenario for these galaxies.

\begin{figure}[h]
   \centering
   \includegraphics[width=9cm]{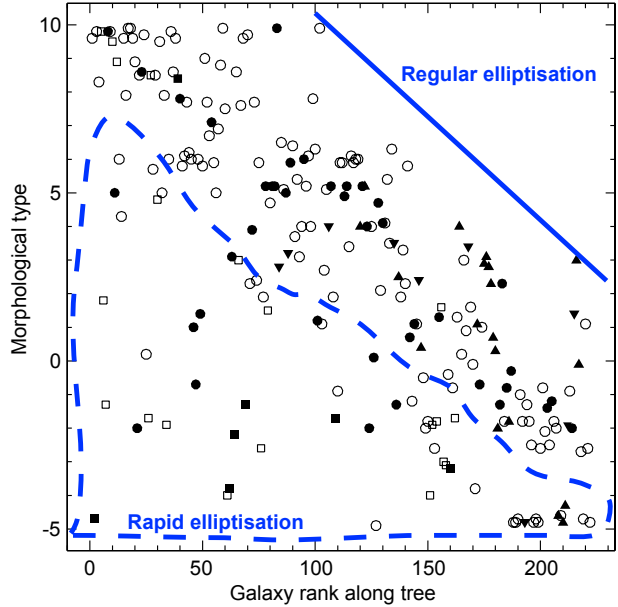}
      \caption{Evolution of the morphology of Virgo galaxies, represented by the de Vaucouleurs index as a function of the rank of galaxies along the tree.}
       \label{figvirgo1}
   \end{figure}
   
One of the most interesting results is that the tree is relatively regular (unbalanced), meaning that the diversification of the Virgo galaxies has been somewhat gentle. The behaviours of all characters confirm that the main drivers of evolution for these objects are soft processes like secular (i.e. isolated) evolution, weak interactions and minor mergers and accretion (see Fraix-Burnet et al. 2006b). 

Several properties indicate that galaxies become bigger and more massive. It is interesting to note that the total amount of HI gas remains nearly constant, with a significant dispersion, while the central surface density of this gas decreases as expected if it forms stars and is then processed. This points toward a continuous income of gas in the outer parts of the galaxies. Also, there is a small evolutionary group of galaxies that is well diversified from the common ancestral group, but shows less massive objects. This could correspond to galaxies that have evolved but have not accreted much matter, maybe because of their particular environment. Galaxy harassment may possibly be invoked as well.

With such an unbalanced cladogram, it is possible to represent the evolutionary scenario by plotting the parameters versus the rank of the galaxies along the tree. For instance, the de Vaucouleurs index of morphology, is plotted in Fig.~\ref{figvirgo1}. Obviously, the Hubble mophological type is not an evolution indicator because no ancestral state can be identified. All types (Irr, S and E) are present at the beginning of the evolution (left side), while there is a clear lack of Irr and S at the end. Hence, we conclude that galaxies mostly become elliptical through evolution. However, since ellipticals are also present on the bottom left of the diagram, the rates for this morphological evolution are diverse. They can be grossly classified into two families: a regular and lengthy process, through accretion, minor mergers or weak interactions, and a more brutal one that corresponds to major mergers and strong interactions. Galaxies all tend to become elliptical, but more or less rapidly depending mainly on their environment. Henceforth, for the morphological parameter, there are several evolutionary paths to achieve the same result.

 \begin{figure}[h]
   \centering
   \includegraphics[width=9cm]{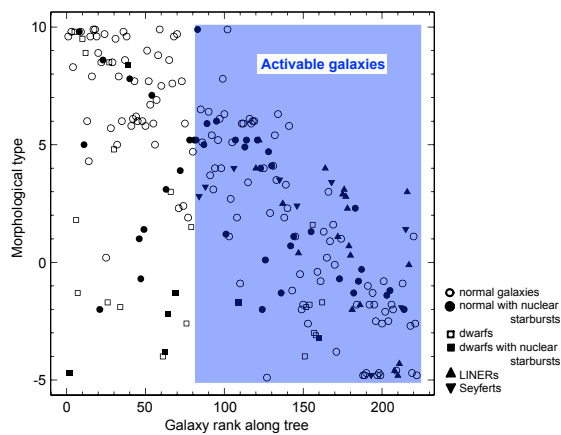}
      \caption{Same as Fig.~\ref{figvirgo1} but with indication of galaxy activity and its onset in the evolutionary scenario. }
       \label{figvirgo2}
   \end{figure}
   
In Fig.~\ref{figvirgo2} we show the same diagram as Fig.~\ref{figvirgo1} but with the shape of the points indicating the activity properties of galaxies, such as nuclear starbursts, LINERs and Seyferts phenomena. It is obvious that the AGN activity appears only on the right side of the diagram, but not in all the corresponding galaxies. We interpret this fact as an evidence that something is acquired by galaxies during the course of evolution, something that allows AGN phenomena and that is transmitted to more evolved kinds of galaxies. This is probably linked to the central black hole but may be not only. We call these galaxies as activable because they form a true evolutionary group, sharing a derived property. In contrast, the nuclear starbursts are present all along the evolutionary history of the Virgo galaxies.

\section{Astrocladistics and big surveys}

Astrocladistics is the only methodology that can fully exploit data of large catalogues and big surveys, by using all available observables at all wavelengths, and by taking into account the evolutionary nature of galaxies. We are finishing the study of about 300 close gas-poor galaxies (Fraix-Burnet \& Davoust in prep.) in different environments. The evolutionary scenario is here more complicated, several evolutionary groups seem to be present. We will use these groups to combine the three samples studied so far, in order to provide an evolutionary scenario of about 500 objects. This work is a prerequisite for our next project to analyse a very large sample of SDSS galaxies. The challenge is important, but the reward can be huge. For the first time, we will be able to depict the diversification of galaxies with a true evolutionary perspective by including objects of the past (at non-zero redshifts).



\end{document}